\documentclass[reprint,superscriptaddress,amssymb]{revtex4-1}
\usepackage{graphicx}
\usepackage[latin1]{inputenc}
\usepackage{color}

\usepackage{times}
\renewcommand{\BibitemShut}[1]{}

\begin{document}

\title{Revisiting the domain model for lithium intercalated graphite }

\author{Sridevi Krishnan}
\affiliation{Laboratoire de simulation atomistique (L\_Sim), SP2M, UMR-E CEA / UJF-Grenoble 1, INAC, Grenoble, F-38054, France}

\author{Gilles Brenet}
\affiliation{Laboratoire de simulation atomistique (L\_Sim), SP2M, UMR-E CEA / UJF-Grenoble 1, INAC, Grenoble, F-38054, France}

\author{Eduardo Machado-Charry} 
\affiliation{Laboratoire de simulation atomistique (L\_Sim), SP2M, UMR-E CEA / UJF-Grenoble 1, INAC, Grenoble, F-38054, France} 
\affiliation{Nanosciences Fondation, 23 rue des martyrs, 38000 Grenoble, France}

\author{Damien Caliste}
\affiliation{Laboratoire de simulation atomistique (L\_Sim), SP2M, UMR-E CEA / UJF-Grenoble 1, INAC, Grenoble, F-38054, France}

\author{Luigi Genovese}
\affiliation{Laboratoire de simulation atomistique (L\_Sim), SP2M, UMR-E CEA / UJF-Grenoble 1, INAC, Grenoble, F-38054, France}

\author{Thierry Deutsch}
\affiliation{Laboratoire de simulation atomistique (L\_Sim), SP2M, UMR-E CEA / UJF-Grenoble 1, INAC, Grenoble, F-38054, France}

\author{Pascal Pochet}
\affiliation{Laboratoire de simulation atomistique (L\_Sim), SP2M, UMR-E CEA / UJF-Grenoble 1, INAC, Grenoble, F-38054, France}


\begin{abstract}

In this letter, we study the stability of the domain model for lithium intercalated graphite in stages \MakeUppercase{\romannumeral 3} and \MakeUppercase{\romannumeral 2} by means of Density Functional Theory and Kinetic Lattice Monte Carlo simulations. We find that the domain model is either thermodynamically or kinetically stable when compared to the standard model in stages \MakeUppercase{\romannumeral 3} and \MakeUppercase{\romannumeral 2}. The existence of domains in the intercalation sequence is well supported by recent high resolution transmission electron microscope observations in lithiated graphite. Moreover, we predict that such domain staging sequences leads to a wide range of diffusivity as reported in experiments.

\end{abstract}


\maketitle

Graphite is an attractive anode material for lithium ion batteries, because of reversible intercalation of lithium with good structural and interfacial stability\cite{Persson2010a}. During the charging cycle, it forms $ Li_{x}C_{6} (0 \leq x \leq 1) $ compounds of well defined stages, with increasing lithium concentration and \emph{vice versa} to form delithiated graphite during the discharging cycle. Stage 'n' corresponds to 'n' number of empty graphene planes sandwiched between two consecutive lithium planes \cite{Dresselhaus2002}. Employing electrochemical and chemical methods \cite{Dahn1991, Billaud1981, Pfluger1981}, the highest observed stage is \MakeUppercase{\romannumeral 4} ($ x \approx 0.2 $) while stage \cite{Guer1975} \MakeUppercase{\romannumeral 1} ($ x = 1 $) is the lowest, with various in-plane lithium ordering. Although, staging and in-plane ordering in these compounds are well established, their structure and the mechanism of staging transition remains unclear. 

Four decades ago, Daumas and Herald \cite{DaumasNHerold1969} proposed the domain model, where the intercalates occupy all galleries as islands. Galleries are the interlayer space between two graphene planes. These islands could be present adjacent to each other in the same gallery, as long as the staging is maintained locally. Consequently, staging transition could occur by exchange of the islands within a gallery, without extensive rearrangement of the intercalates. 
Several theoretical studies \cite{SafranSA1980, DDL1987} support the domain model and explain the staging mechanism considering elastic and electrostatic interactions. Treating the graphite layers as elastic plates, it was found that the intercalants within a gallery form islands by attracting each other\cite{KirczenowG1982}  but beyond a critical size they have a barrier to merge. The islands in the neighboring galleries attract each other favoring staggered domains.\cite{Hamann1979} Assuming well defined domains, the staging transition is proposed to occur by rotation of domain walls\cite{ForgacsGUiman1984} and through order-disorder transitions.\cite{KirczenowG1985} In addition, the nucleation, growth and merging of small elementary islands to form larger ones, during stage transition\cite{Kirczenow1985} are also predicted. These studies provide a base for understanding the domain model and intercalation mechanism in these compounds, but, the specific case of lithium intercalant in the domain model is yet to be addressed at the Density Functional Theory (DFT) level.    

On the other hand, using \emph{ab-initio} methods, several authors\cite{Kganyago2003, N.A.W.HolzwarthLouieStevenG1981, Imai2007, Woo1983a, Review1984, Persson2010, Qi2010} have studied the stability of different stages of the lithium intercalated graphite in the experimentally observed stoichiometric structures and in some hypothetical structures. These calculations consider a periodic unit cell with infinite sheets of intercalates in the galleries (referred as standard (std.) model hereafter). This is adequate from a thermodynamic perspective, but from a kinetic point of view, it is insufficient to explain the staging transition. Here, the staging transition proceeds by Li diffusion either across the graphite layers or migration around them \cite{Dresselhaus2002}. The diffusion of lithium through the basal plane is highly unlikely, because of the high barrier for migration $ \approx $ 10 eV. Likewise, migration around the graphite layers can happen only at the edges and boundaries. Within this representation, entire galleries have to be completely emptied and others have to be completely filled with lithium for staging transition to occur. It is specially complicated for a stage transition from an even indexed stage to an odd one or \emph {vice versa}. 
 
The domain model for lithium intercalated graphite serves as a reasonable model to explain the staging behavior of these compounds. Considering the ubiquitous presence of graphite anodes in most commercial Li-ion batteries, the first principles study of this model will give us important insights about the structure of the anode and the staging mechanism involved. Here, we study the domain structure for lithium intercalated graphite in stages \MakeUppercase{\romannumeral 3} and \MakeUppercase{\romannumeral 2} based on the domain model using DFT and compare their stability with the standard model. The stability of this compound in the domain model has potential implications for lithium diffusion in graphite.

All \textit{ab-initio} calculations are performed within the DFT approach as implemented in BigDFT\cite{LuigiBig2008} code based on the wavelet basis set. The Perdew-Burke-Ernzerhof\cite{PBE} (PBE) functional  was used for approximating the exchange correlation employing the Hartwigsen-Goedecker-Hutter\cite{HGH} pseudo potentials with the Krack variant.\cite{krack} The number of basis functions and kpoints were chosen to obtain an accuracy of 1 meV/atom. The corresponding grid spacing for the uniform wavelet basis is 0.36 bohr. Atomic relaxation was carried out until the forces acting on each atom was less than 0.02 eV/{\AA}. To account for the inter planar Van der Waals (VdW) interactions between the graphene planes, we use a dispersion correcting scheme, Dispersion Corrected Atom Centered Pseudopotentials \cite{VonLilienfeld2004} (DCACP) on the PBE functionals. DCACP accounts for the long-range electron correlation by using potentials that have been calibrated against benzene dimer as reference system for carbon.  

Grand canonical Kinetic Lattice Monte Carlo (KLMC) simulations have been performed to simulate stage transitions. The lattice used for this simulation considers only the Li sites located above the hexagon center of a carbon sheet in a $ LiC_{6} $ lattice. In addition, only one hexagon center site for Li is kept for every three to ensure that the $\sqrt{3}\times\sqrt{3}$ arrangement of lithium is preserved over the simulation box. The simulation cell corresponds to a $12\times8\times6$ orthorhombic $ LiC_{6} $ cell with 1152 Li out of a total of 8064 atoms. A Li reservoir is fixed at $ x=0 $ surface of the cell while the other directions are kept periodic. The underlying energetic model will be described later in the discussion part.

We first compare the deviation of the calculated lattice parameters (std. model) from the experimentally reported values in the $ Li_{x}C_{6} $ compounds, for $ x =$ 0, 0.33, 0.5 \& 1 corresponding to graphite, stage \MakeUppercase{\romannumeral 3}, stage \MakeUppercase{\romannumeral 2} and stage \MakeUppercase{\romannumeral 1}. The in-plane lattice constant is predicted very well with a deviation of less than 1\% in all cases. However, the deviations are larger for the interlayer lattice parameter. The PBE functional does not account for the long range interaction and shows no inter-planar binding in graphite and the deviation for  $ x = $ 0.33 and 0.5 is $ \approx $ 7 \% and $ \approx $ 6\% respectively. On the other hand, this functional performs well (\textless 1\% deviation) for $ x = $ 1, which is mostly ionic due to the lithiation of all galleries. To avoid spurious Li-C interactions and the resulting over-binding in lithiated galleries, we have mixed the DCACP and PBE pseudopotentials  for some carbon atoms chosen depending on their type of interlayer interactions i.e., either VdWs or ionic interactions. The mixed DCACP shows improved performance for systems with VdW contributions and the deviation is $ \approx $ 2\% , $ \approx $ 5 \% and $ \approx $ -2\%  for graphite and for $ x = $0.33 and 0.5 respectively. We use the mixed DCACP potentials for all further studies on the domain model.

\begin{figure}
\includegraphics[width=0.5\textwidth]{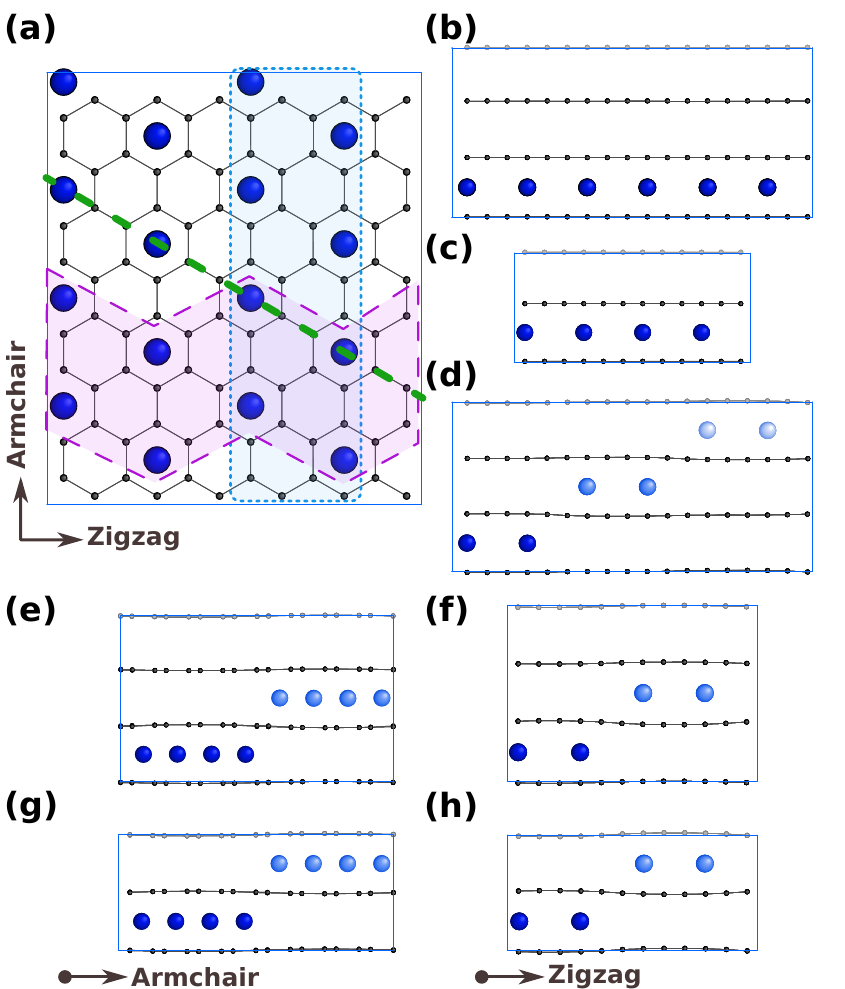}
\caption{(color online) Optimized structures of $ Li_{x}C_{6} $ compounds. The black and blue spheres represent C and Li atoms respectively. (a) Top view of the $ \sqrt{3} \times \sqrt{3} $ arrangement of Li atoms in graphite lattice. The dotted (blue) area refers to the straight (STR) edged strips, periodic in armchair (AC) direction and the dashed (pink) area refers to the zigzag (ZZ) edged strips, periodic ZZ direction. (b-c) The standard model for stages \MakeUppercase{\romannumeral 3} and \MakeUppercase{\romannumeral 2} respectively. (d) Domain models with a width of 2 Li strips for Stage \MakeUppercase{\romannumeral 3} in STR strips domains. (e-f) Domain models with a width of 2 Li strips occupying only 2/3\textsuperscript{rd} of the galleries, for Stage \MakeUppercase{\romannumeral 3} in ZZ and STR edged strips respectively. (g-h) Domain model for stage \MakeUppercase{\romannumeral 2} in ZZ and STR edges respectively. The different shades of blue spheres represent Li atoms in different layers. The darker the spheres the deeper the intercalant. The green (thick) dashed line in (a) corresponds to the plane through which charge density difference maps are shown in figure ~\ref{fig:DHchargeden}} 
\label{fig:DHmodels}
\end{figure}

The in-plane domains are best modeled as islands of different shapes and sizes, however, it is computationally expensive. In addition, the size will increase significantly with increasing domain size and stage index. We propose restricting the periodicity of intercalates in one dimension to reduce the size of supercell. Accordingly, we study 1-D strip domains periodic either in zigzag (ZZ) or armchair (AC) direction,  forming ZZ or straight (STR) edges (see figure.~\ref{fig:DHmodels}(a)). For comparison, a 0-D in-plane island was constructed and studied for stage \MakeUppercase{\romannumeral 2}. Since, lithium occupies all galleries in this model, the stacking sequence of the graphene sheets is always kept at A$ \alpha $A$ \alpha $A. The supercell for stage \MakeUppercase{\romannumeral 3} consisted a total of 304 atoms for when the galleries were 2/3\textsuperscript{rd} filled and 456 atoms when completely filled . For stage \MakeUppercase{\romannumeral 2} it was 208, 312 and 416 atoms for systems with a width of 2, 3 and 4 number of Li strips.

The optimized structures in both the stages are shown in figure.~\ref{fig:DHmodels}(b-h). The graphene sheet is corrugated along the ZZ or AC directions for STR or ZZ edged strips. The graphene layer deforms on either side of the strip forming the boundary region between successive domains and therefore, the corrugation wavelength depends on strip width. In addition, we find that the corrugation amplitude is higher in case of STR strips compared to the ZZ edges due to the nature of the strip edges. The interlayer distance, the lithium-lithium separation and the Li-C bond length are strained in these structures near the domain boundary. Compressive strain of up to 2\% in Li-Li separation are observed for the smallest strip size (2 Li strips) in stage \MakeUppercase{\romannumeral 2} (fig. ~\ref{fig:DHmodels}(h)). Since, the domain size is inversely proportional to stage index\cite{Dresselhaus2002}, the size of the domains in stage \MakeUppercase{\romannumeral 2} should be larger than that in stage \MakeUppercase{\romannumeral 3}, accordingly we increase the number of strips in each gallery for this stage and find that the deformation reduces to \textless 1\% for the largest strip size (width of four Li strips). 

To compare the stability of the domain model with that of their standard counterparts, we study their formation energy. (see Table ~\ref{tab:DHform}). The formation energy per formula unit is calculated as $ E_{f} = (E(Li_{x}C_{6}) - x\mu_{Li}^{\tiny metallic} - 6\mu_{graphite})/6 $. The stage \MakeUppercase{\romannumeral 3} compound in the smallest domain size with two Li-strips in all galleries is more stable by 8.6 meV (figure.~\ref{fig:DHmodels}(d)), compared to the std. model. The same is true for the system with two strips in two third of the galleries in stage \MakeUppercase{\romannumeral 3} (figure.~\ref{fig:DHmodels}(e \& f)). Interestingly, altering the edge type has negligible effect on the stability of the compound. The stage \MakeUppercase{\romannumeral 2} compounds are comparable in energies and differ only by a few meVs. Nevertheless, the difference in formation energy of the domain model and the std. model decreases with increasing strip size in accordance with the reduced in-plane compressive strain. In addition, the formation energies of the island model in stage \MakeUppercase{\romannumeral 2} is close to (-20.6 meV/C atom) the corresponding strip model which validates our use of the strip model.  

\begin{table}[h]
\begin{tabular}{ccccc}
\hline\hline
 &  & \multicolumn{3}{c}{Formation energy ($ E_{f} $) (meV)} \\
 \cline{3-5}
& &  &\multicolumn{2}{c}{Domain model} \\ 
 \cline{4-5}
Stage& \# Li strips  & Std. &  STR edge &  ZZ edge \\ 
\hline
\MakeUppercase{\romannumeral 3}   & 2  & -9.5 & -18.1 & \\
                                  & 2-2/3\textsuperscript{rd}& -9.5 & -16.3 & -17.0 \\
\hline
\MakeUppercase{\romannumeral 2}   & 2 & -22.9 & -20.7 & -21.3 \\
                                  & 3 & -23.1 & -21.0 &   \\
                                  & 4 & -23.6 & -22.9 &   \\
\hline\hline
\end{tabular}
\caption{Formation energy of the strip domain model with increasing strip size along with the formation energy of the corresponding standard model. }
\label{tab:DHform}
\end{table}

To understand the rationale behind the stabilization of the different stages we analyze the different contributions to these compounds in detail. The thermodynamic stability of these structures originates from the balance between the energetic contributions from elastic and electrostatic interactions. We observe that the elastic contribution from the deformation in the graphene sheet plays a minor role in the stabilization of the compounds. It was calculated from single point energies of the graphite deformed after lithium intercalation in the domain model. Indeed, the elastic energy of the deformed graphite revealed that it remained unchanged with deformation and increasing strip size for stage \MakeUppercase{\romannumeral 2}. In addition, for stage \MakeUppercase{\romannumeral 3}, the elastic contribution was less than +2 meV/C atom. Thus the elastic energy due to graphene corrugations does not explain the stability of these compounds.

\begin{figure}
\includegraphics[width=0.5\textwidth]{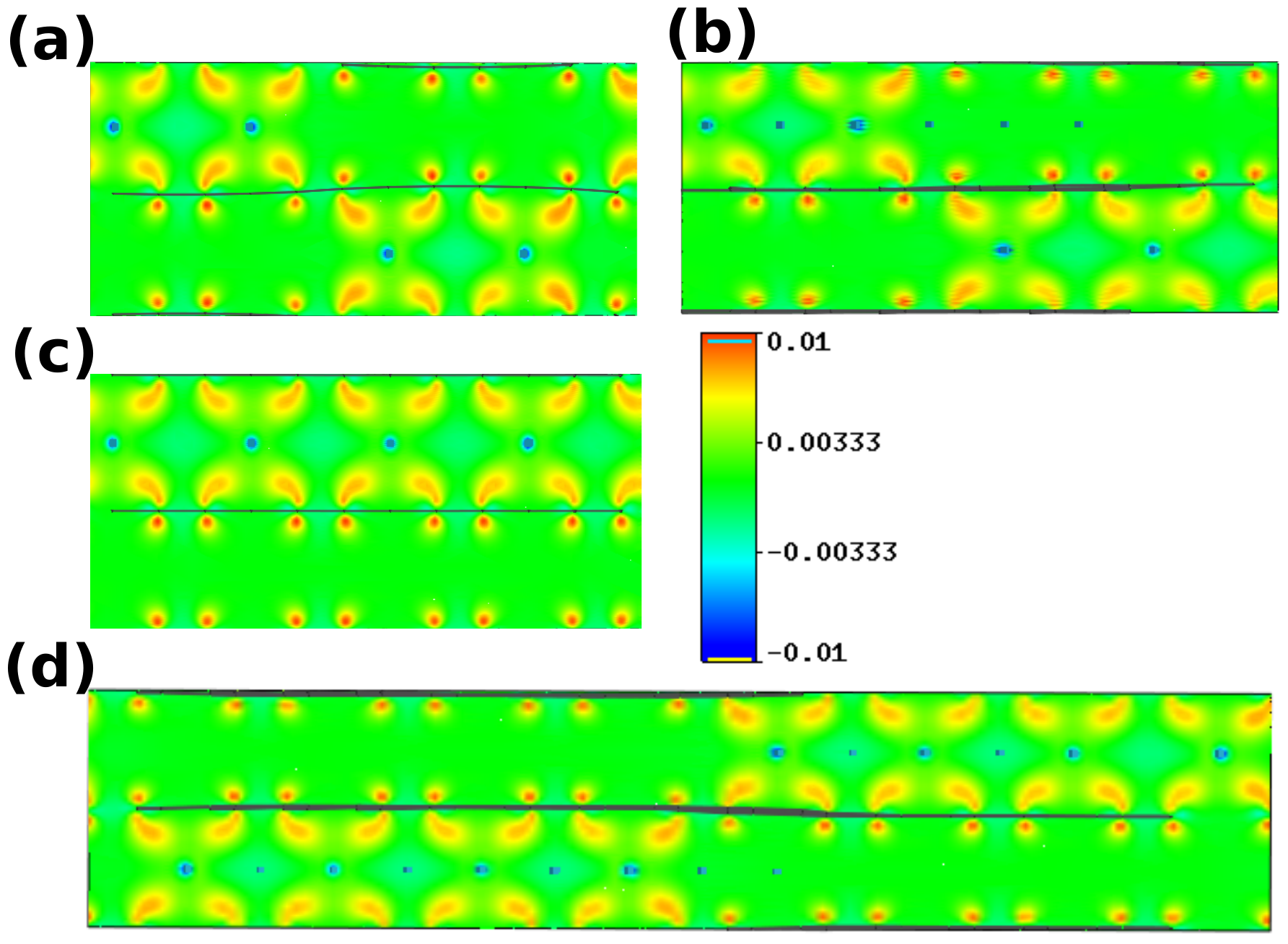}
\caption{(color online) Charge density difference map projected to the plane marked by a green (thick) dashed line in fig.~\ref{fig:DHmodels}(a) for Stage \MakeUppercase{\romannumeral 2}. (a), (b) \& (c) 2 Li strips for STR edged domain, ZZ edged domain and standard model respectively. (d) 4 Li strips for STR edged strip model.}
\label{fig:DHchargeden}
\end{figure}

Therefore, we analyze the electrostatic contributions to explain the stability of these compound. Bader charge analysis shows that lithium atoms transfer its charge to the carbon atoms, remaining in an ionic state in the compound. The charge on the Li and C atoms are found to be +0.86e and -0.07e respectively in stages  \MakeUppercase{\romannumeral 3} and  \MakeUppercase{\romannumeral 2}. The C atoms with no Li neighbors in stage \MakeUppercase{\romannumeral 3} are neutral. To study the inter-plane interaction between Li and C we compare the charge density difference maps of the stage \MakeUppercase{\romannumeral 2} compounds. The charge density difference was calculated as $ \Delta\rho(r) = \rho_{\tiny Li_{x}C_{6}}(r) - \rho_{\tiny Li}(r) - \rho_{\tiny C}(r)$, where $ \rho_{\tiny Li_{x}C_{6}}(r) , \rho_{\tiny Li}(r) $ and  $\rho_{\tiny C}(r)$  are the charge densities of the $ Li_{x}C_{6} $ compounds, Li in graphite host and that of graphite without the Li charge densities respectively. We identify two types of lithium atoms either with symmetric or asymmetric charge distribution surrounding them. For a standard model the surroundings of the Li atoms is unchanged and are always symmetric (fig ~\ref{fig:DHchargeden}(c)). But for the domain model, the ones in the domain center are symmetric while those in the domain boundary are asymmetric (Fig ~\ref{fig:DHchargeden}(a, b \& d)). The ratio of symmetric to asymmetric Li atoms changes with strip width. While for the 2 Li width model all Li atoms are asymmetric, for the 4 Li width model, there are both types of atoms. With increasing strip size, the number of symmetric lithium atoms increases, following the change in formation energies in these compounds. We also observe that the change in charge transfered to the C atoms at the boundary, which are bound to a Li atom with asymmetric charge distribution is insignificant. Thus, this provides evidence that the electrostatic interactions dominate the stability in these models.

So far, the energetics demonstrate the thermodynamic stability of the domain model only for stage \MakeUppercase{\romannumeral 3}. However, the kinetics could play a vital role in the stage transformation and alter the Li intercalation. We therefore employ KLMC method to simulate stage \MakeUppercase{\romannumeral 3} to \MakeUppercase{\romannumeral 2} transition starting from both the standard and the strip model. We choose an energetic model that favors standard staging arrangements over disordered lithium or even domain arrangements. The energies per lithium are 0.85 eV, 0.71 eV and 0.63 eV in standard stages \MakeUppercase{\romannumeral 1}, \MakeUppercase{\romannumeral 2} and \MakeUppercase{\romannumeral 3} respectively. These energies are modified by +0.14eV and +0.11eV (for stages \MakeUppercase{\romannumeral 2} and \MakeUppercase{\romannumeral 3} respectively) when a lithium is located at a domain step. For kinetic evolution, we choose a barrier of 0.24~eV considering a unique jump to reach the second neighbor hexagonal Li site in the gallery. It is based on the diffusion barrier of 0.20 eV \cite{Persson2010} for the first neighbor jump plus an analytical correction to reach the second neighbor in one jump. A higher barrier (0.40 eV\cite{Persson2010}) in the presence of a neighboring Li prevents migration when the first neighbor is occupied. The simulated time lasts for $\approx$1$\mu$s ($ 8 \times 10^{7} $ steps) at 500K and with a Li chemical potential of 0.90 eV. The elevated temperature is chosen to accelerate the convergence without altering the energy landscape. The choice of the chemical potential value is governed by the incorporation kinetics. A lower value would make the incorporation very slow and difficult to simulate, while a higher value would correspond to a fast charge, filling up the graphite up to stage \MakeUppercase{\romannumeral 1} without any intermediate stages.

\begin{figure}
\includegraphics[width=0.5\textwidth]{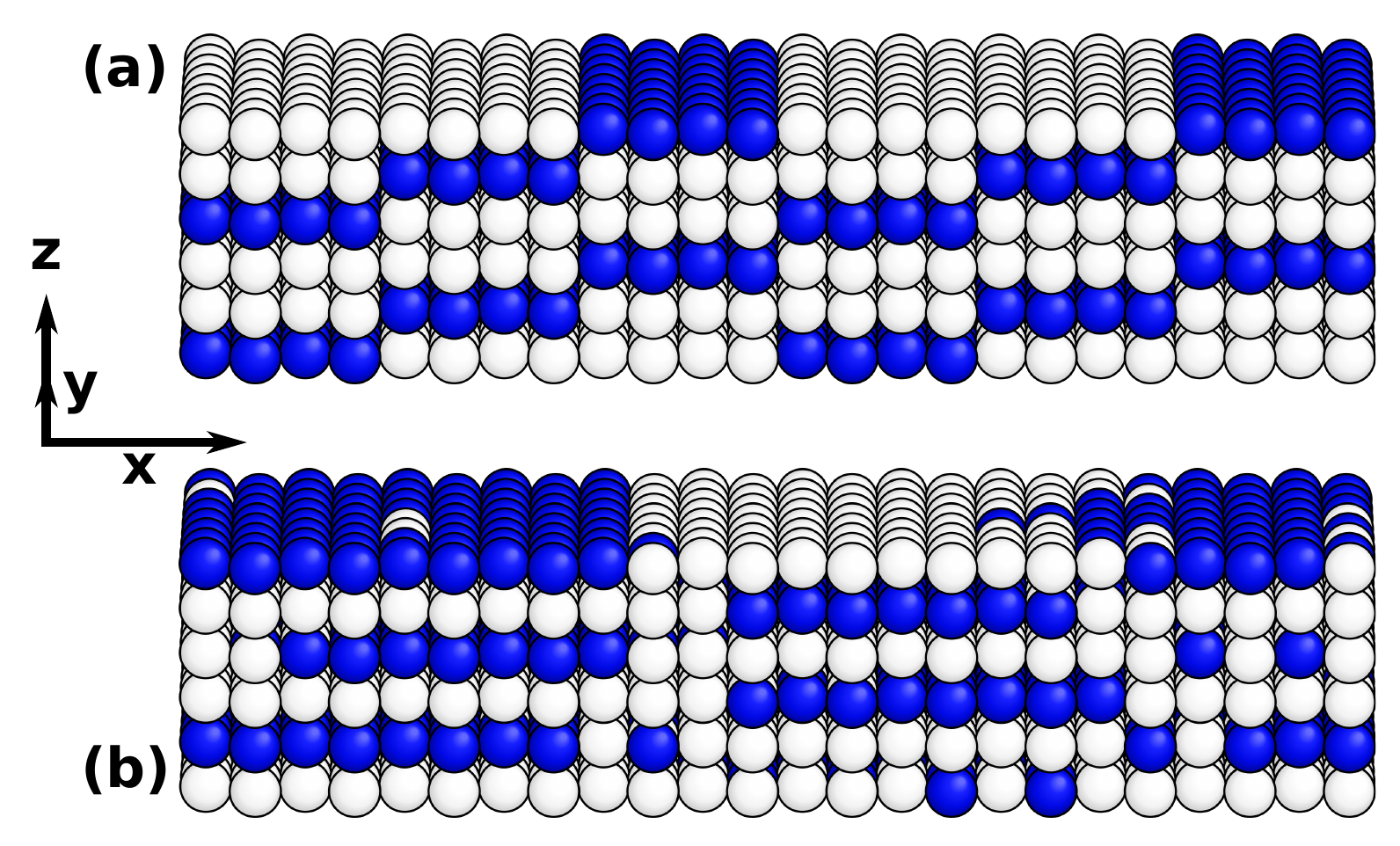}
\caption{\label{fig:Li_gr_diff} (color online) (a) The initial configuration of the Monte Carlo simulation cell in stage \MakeUppercase{\romannumeral 3} in strip model and (b) the evolution of the former into stage \MakeUppercase{\romannumeral 2} after $\approx$1$\mu$s. Blue and white spheres represent the Li atoms and vacancies respectively. The interface with the reservoir is at $ x=0 $. }
\end{figure}

We find that a standard stage \MakeUppercase{\romannumeral 3} model cannot evolve within the simulation time. However, starting from a strip model presents a quicker kinetic evolution and it transformed from a domain stage \MakeUppercase{\romannumeral 3} into a domain stage \MakeUppercase{\romannumeral 2}. The initial and the final cell configurations for the strip model are shown in Fig. ~\ref{fig:Li_gr_diff}.  It is clear that the strips have reorganized and the strip size has increased in stage \MakeUppercase{\romannumeral 2}. In conjunction with our \textit{ab-initio} results where standard stage \MakeUppercase{\romannumeral 2} has a greater stability over smaller domains, these KLMC results are in favor of an intercalating scenario where the graphite is lithiated by domains and the domain size grows with the concentration of lithium. Another interesting observation is that the electrode edge ($ x=0 $) presents an almost perfect stage \MakeUppercase{\romannumeral 2} whereas the opposite $x$ edge resembles the center of the electrode with scare Li concentration. This result is consistent with  the recent experimental evidence of gradient concentration in the electrode (see Fig.~4 $b$ of ref.~\cite{Wang2011}) and of good stage ordering near the interface (same reference, Fig.~4 $c$).

Thus far, we have reported DFT and KLMC simulations that supports the intercalation sequence through domains. Such a sequence is found to be in line with recent transmission electron microscopy observations \cite{Wang2011} that reports spatial non-uniformity in the  lithium intercalation and concentration. Moreover, our results indicate that though the elastic energy contributes a major part to the system, the electrostatic energy governs the stability of the compounds over one another. 
\textbf{Furthermore, the corrugations introduced in the graphene sheets in domain model results in varying interlayer distance within a single gallery.
Such a variation provides an alternative explanation for the recently reported stacking disorder in stage  \MakeUppercase{\romannumeral 3} \cite{Rune2013}.
Besides, it has }been shown\cite{Xu2012a} that the migration barrier for lithium in graphite is strongly correlated with interlayer spacing. 
Accordingly, a lithium atom near the domain boundary has a higher barrier compared to one in the center of the domain. For instance, in our case the change in interlayer distance within a single gallery ranges from 3.4 \AA\ to 3.8 \AA\ , the change in barrier reported\cite{Xu2012a} for this range is $\sim$ 0.2 eV. Using these barriers and calculating the chemical diffusivity at 300K with $\nu^{*} = 1\times10^{13}s^{-1}$, \cite{Toyoura2008} the change in diffusivity within a single gallery is found to vary from $ \sim 10^{-9}$ to $\sim 10^{-12}$ $cm^2/s$. 
Our results then offer an alternative explanation to the reported range of diffusivity \cite{Persson2010a} and indicates that even with a rational design of the electrodes, 
the formation of domains leads to changing diffusivity within a single gallery. 
Detailed experiments on the diffusivity of Li at different stages and lithium concentration would be needed to confirm our prediction.   

In summary, the domain model proposed for graphite intercalation compounds has been studied in detail for stages \MakeUppercase{\romannumeral 3} and \MakeUppercase{\romannumeral 2} using atomistic simulations. 
Our DFT results favor the thermodynamic stability of the strip domain model in stage \MakeUppercase{\romannumeral 3} and shows comparable energies for stage \MakeUppercase{\romannumeral 2} independent of the type of domain edge. 
We also observe that the electrostatic interactions play a major role in the stabilization of these compounds. 
Grand canonical KLMC simulation of stage transitions from stage \MakeUppercase{\romannumeral 3} to \MakeUppercase{\romannumeral 2} reveals that the domain model is favored from a kinetic point of view. 
In addition, the transition to stage \MakeUppercase{\romannumeral 2} is accompanied by an increase in strip size of the domains. 
These results favor lithiation of graphite in the domain structure and increasing domain size with the lithium concentration. The stability of the domain model is in agreement with recent experimental observations and our prediction offers an alternative explanation for the range of diffusivities observed in experiments.
Our findings clearly show that kinetics dictates the staging transition in these compounds which in turn can occur only with the domain model. 

This work was performed using computational resources (GENCI-2013096705 and IDRIS-GC-2012). 
SK was supported by the NTE CEA-program.
PP acknowledges the  EC FP7 projects MMM$@$HPC (Grant no.261594) and Hi-C (Grant no. 608575) for partial funding.

\end{document}